\documentstyle[e-e-ijmpa,epsfig,citepunct]{article}

\def\ARNP#1#2#3{Ann. Rev. Nucl. Part. Sci. {\bf#1}, #2 (#3)}

\def\IJMP#1#2#3{Int. J. Mod. Phys. {\bf A#1}, #2 (#3)}

\def\NPB#1#2#3{Nucl. Phys. {\bf B#1}, #2 (#3)}

\def\PLB#1#2#3{Phys. Lett. {\bf B#1}, #2 (#3)}

\def\PRep#1#2#3{Phys. Rep. {\bf#1}, #2 (#3)}
\def\PRD#1#2#3{Phys. Rev. {\bf D#1}, #2 (#3)}
\def\PRL#1#2#3{Phys. Rev. Lett. {\bf#1}, #2 (#3)}

\def\ZPC#1#2#3{Zeit. f\"ur Physik {\bf C#1}, #2 (#3)}

\newcommand{\lesssim}{\mathop{}_{\textstyle \sim}^{\textstyle <}}
  
\newcommand{\emem}{e^-e^-}
\newcommand{\epem}{e^+e^-}
\newcommand{\ifb}{~{\rm fb}^{-1}}
\newcommand{\yr}{~{\rm yr}}
\newcommand{\mev}{~{\rm MeV}}
\newcommand{\gev}{~{\rm GeV}}
\newcommand{\tev}{~{\rm TeV}}

\newcommand{\selectron}{\tilde{e}}
\newcommand{\text}[1]{ {\rm #1} }
\newcommand{\itemjlf}[1]{\item {\em #1}}
\newenvironment{mylist}{\begin{list}{$\bullet$ }
{\addtolength{\leftmargin}{-0.2in}\rightmargin=\leftmargin
\setlength{\itemsep}{-0.01in}
\setlength{\labelsep}{0.0in}}
\vspace*{-.05in}}
{\end{list}}

\renewcommand{\thefootnote}{\fnsymbol{footnote}} 

\begin{document}

\normalsize\textlineskip

\noindent
\begin{minipage}[t]{\textwidth}
\begin{flushright}
hep-ph/0002055 \hfill
IASSNS--HEP--00--10\\
\end{flushright}
\end{minipage}
\vspace*{.3in}

\title{\LARGE \bf
Physics at \boldmath$\emem$ Colliders
\vspace*{.3in}
}

\author{\large Jonathan L. Feng
}

\address{\normalsize
\vspace*{0.2in}
School of Natural Sciences, Institute for Advanced Study\\
Princeton, New Jersey 08540 \, USA}

\maketitle

\vspace*{1.2in}

\abstracts{\normalsize 
\centerline{An overview of the physics motivations for $\emem$
colliders is presented.} 
{\normalsize
\vspace*{2.0in}
\begin{center}
Opening lecture of the 3nd International Workshop on\\
Electron-Electron Interactions at TeV Energies ($\emem$99)\\
University of California, Santa Cruz, 10--12 December 1999
\end{center}}
}

\setcounter{footnote}{0}
\renewcommand{\thefootnote}{\alph{footnote}}

\vspace*{1pt}\textlineskip	

\newpage

\section{Introduction}

The goal of this meeting is to encourage the exchange of new ideas
about TeV-scale $\emem$ colliders, as well as their $e^-\gamma$ and
$\gamma\gamma$ cousins.  Such colliders are to be considered as a
component of future linear collider programs along with the next
generation of $\epem$ colliders.  At first sight, the $\emem$ option
might appear to require only trivial modifications of the $\epem$
mode.  In fact, many interesting issues arise if one wants to optimize
the $\emem$ performance to, for example, obtain luminosities
comparable to those in the $\epem$ mode. These issues have been
addressed previously~\cite{95,97} and will also be discussed at this
meeting.  Nevertheless, there is broad consensus that, with planning,
the $\emem$ mode is a relatively simple, inexpensive, and
straightforward addition to any linear collider program.

There are many thorny questions regarding when, where, and how such
colliders should be funded and built --- these issues are far beyond
the scope of this talk.  Rather, I will address a more modest (and
much more intriguing) question: what novel and exciting possibilities
for exploring weak scale physics will an $\emem$ collider provide?

\section{New Physics}

The standard model is now verified to extraordinary
accuracy~\cite{Quigg}.  The strong, weak, and electromagnetic gauge
couplings have been determined through numerous independent
measurements and are known to 1 part in $10^{2}$, $10^{3}$, and
$10^{8}$, respectively.  In the matter sector, there are now three
complete generations with, for the most part, well-known masses and
mixings.  Even ten years ago, despite an intervening decade typically
regarded as ``quiet,'' this story would have been far less complete.
The contributions of the SLC, LEP, Tevatron, and HERA colliders at the
high energy frontier have done an impressive job of bringing the
present picture into sharp focus.

Sharp focus often leads to a greater appreciation of blemishes,
however, and this is the case with the standard model.  Some of the
outstanding puzzles are the problems of

\begin{mylist}

\itemjlf{Electroweak Symmetry Breaking.} We do not understand the
fundamental mechanism of electroweak symmetry breaking and the source
of the gauge hierarchy, despite (or, better, as demonstrated by) the
existence of many proposed solutions.

\itemjlf{Flavor.} An explanation of the masses, mixings, and CP
violation observed in the fermion sector remains a complete mystery,
despite an abundance of data.

\itemjlf{Gravity and Spacetime Structure.}  Our understanding of
gravity is limited, and the spacetime structure of our universe is
open to wild speculation. It is remarkable that one can place two
fingers slightly less than a millimeter apart and not know whether
their interaction is primarily gravitational. The problem of the
cosmological constant is emblematic of the lack of understanding in
this area.  It is tempting to speculate that, as we enter the 21st
century, the cosmological constant problem is a hint of fundamental
change on the horizon, just as black body radiation was in the
previous century.  Certainly we should not exclude such a possibility.

\end{mylist}

Given all these mysteries, what are the lessons for future colliders?
The standard approach is to examine the prospects for a given collider
to probe a simple realization of one theoretical idea, and then
another, and another, etc.  Without actual data to guide us, this is
probably the best we can do, and it will be the approach taken below.
Before doing so, however, let us remind ourselves of the following
caveats:

\begin{mylist}

\itemjlf{New physics may be complicated.}  Studies of new physics
typically consider some simple prototypes that are hoped to capture a
few essential features.  To give a concrete example, in supersymmetry,
studies are often done in some minimal framework with few parameters.
It is highly unlikely that such prototypes will be realized in nature.
(It would also be truly disappointing if they were, as these
prototypes are typically based on completely {\em ad hoc} assumptions,
and the consistency of nature with such bland and unmotivated models
would probably leave us at a loss for suggestive clues pointing toward
further progress.)

\itemjlf{New physics need not be modular.}  It is an obvious
possibility that several different types of new physics may reveal
themselves simultaneously, considerably complicating their
interpretation.  One need only look at the last two chapters in the
story of charged lepton discovery (the $\mu$--$\pi$ and $\tau$--charm
puzzles) to find historical precedents.  Again taking supersymmetry as
an example, additional gauge bosons~\cite{Z'} and extended Higgs
sectors are just some of the many possible extensions beyond the
minimal supersymmetric standard model.

\itemjlf{New physics need not appear in its entirety.}  For example,
in strongly coupled theories, only part of a resonance may appear, or
in extra dimensional scenarios, perhaps only one Kaluza-Klein mode
will be unveiled.  Similarly, only a small fraction of the
supersymmetric spectrum is required by naturalness to be at the weak
scale.

\end{mylist}

Of course, it is possible that future colliders will discover only a
standard model-like Higgs boson.  It is also possible, however, that
they may uncover so much anomalous data that it will be decades before
a new synthesis is achieved.  Given the number of fascinating
fundamental questions remaining, some of which are intimately tied to
the weak scale, I find the latter possibility far more likely.

\section{Unique Features of \boldmath$\emem$ Colliders}

If anything like the scenario just described is realized, it is clear
that the future will require a flexible high energy physics program to
make many model-independent measurements.  With the LHC, $\epem$
colliders go a long way toward realizing this goal.  Such colliders,
with specifications
\begin{eqnarray}
\sqrt{s} &=& 0.5 - 1.5 \tev \nonumber \\
{\cal L} &=& 50 - 500 ~\ifb/\yr \qquad (1 \ R = 10^4 - 10^5
~\text{events/yr}) \nonumber \\
P_{e^-} &\equiv& \frac{N_R-N_L}{N_R+N_L} \simeq 90\% 
\qquad (\Delta P_{e^-} \lesssim 1\% )  \ ,
\end{eqnarray}
where $P_{e^-}$ is the electron beam polarization, have been studied
extensively.  While their virtues and drawbacks can only be defined
precisely on a case-by-case basis, it is possible to come to some
general conclusions.  The most salient virtues of $\epem$ colliders
have been summarized by Murayama and Peskin~\cite{MP} as

\begin{mylist}

\itemjlf{Holism.}  At $\epem$ colliders, complete events yield more
information than the sum of their parts.  In other words, the
well-specified initial energy and initial state $\epem_{L,R}$ yield
important constraints.

\itemjlf{Cleanliness.}  Backgrounds are small, and may be reduced with
beam polarization in many cases.

\itemjlf{Democracy.}  The $\epem$ initial state is electrically
neutral and has no overall quantum numbers.  Thus, both lepton and
hadronic sectors may be explored with comparable statistics.

\end{mylist}

Following this rubric, let us now consider the properties of $\emem$
colliders:

\begin{mylist}

\itemjlf{Extreme Holism.}  At $\emem$ colliders, the initial energy is
again well known, but now the initial state may, in principle, be {\em
exactly} specified by the possibility of highly polarizing both beams.

\itemjlf{Extreme Cleanliness.}  Backgrounds are typically extremely
suppressed, and are even more readily reduced by the specification of
both beam polarizations.

\itemjlf{Dictatorship of Leptons.}  Here $\emem$ and $\epem$ colliders
differ sharply: in $\emem$ mode, the initial state has electric charge
$Q=-2$ and lepton number $L=2$.

\end{mylist}

With respect to the first two properties, the $\emem$ collider takes
the linear collider concept to its logical end.  The third property
makes $\emem$ colliders unsuitable as general purpose colliders, but,
as we will see, it is also the source of many advantages.

\section{Case Studies}

There are many interesting opportunities for $\emem$, $e^-\gamma$, and
$\gamma\gamma$ colliders to probe new physics.  I will highlight a few
examples that illustrate the general remarks above.

\subsection{M{\o}ller Scattering}

The process $\emem \to \emem$ is, of course, present in the standard
model.  At $\emem$ colliders, the ability to polarize both beams makes
it possible to exploit this process fully.

For example, one can define two left-right asymmetries
\begin{eqnarray}
A_{LR}^{(1)} &\equiv& 
\frac{d\sigma_{LL} + d\sigma_{LR} - d\sigma_{RL} - d\sigma_{RR}}
     {d\sigma_{LL} + d\sigma_{LR} + d\sigma_{RL} + d\sigma_{RR}}
\nonumber \\
A_{LR}^{(2)} &\equiv& 
\frac{d\sigma_{LL} - d\sigma_{RR}}
     {d\sigma_{LL} + d\sigma_{RR}} \ ,
\end{eqnarray}
where $d\sigma_{ij}$ is the differential cross section for $e^-_i
e^-_j \to e^- e^-$ scattering.  There are four possible beam
polarization configurations.  Assume that the polarizations are
flipped on small time intervals.  The number of events in each of the
four configurations, $N_{ij}$, depends on the two beam polarizations
$P_1$ and $P_2$.  If one assumes the standard model value for
$A_{LR}^{(1)}$, the values of $N_{ij}$ allow one to simultaneously
determine both $P_1$ and $P_2$ (and also $A_{LR}^{(2)}$).  For
polarizations $P_1 \simeq P_2 \simeq 90\%$, integrated luminosity $10
\ifb$, and $\sqrt{s}=500\gev$, Cuypers and Gambino have shown that the
beam polarizations may be determined to $\Delta P / P \approx
1\%$~\cite{CB}.  Such a measurement is comparable to precisions
achieved with Compton polarimetry, and has the advantage that it is a
direct measurement of beam polarization at the interaction point.

Perhaps even more exciting, this analysis also yields a determination
of $A_{LR}^{(2)}$, as noted above.  Any inconsistency with the
standard model prediction is then a signal of new physics.  For
example, one might consider the possibility of electron compositeness,
parameterized by the dimension six operator ${\cal L}_{\rm eff} =
\frac{2\pi}{\Lambda^2} \bar{e}_L \gamma^{\mu} e_L \bar{e}_L
\gamma_{\mu} e_L$.  Barklow has shown that with $\sqrt{s} = 1 \tev$
and an $82 \ifb$ event sample, an $\emem$ collider is sensitive to
scales as high as $\Lambda = 150 \tev$~\cite{Barklow}.  The analogous
result using Bhabha scattering at $\epem$ colliders with equivalent
luminosity is roughly $\Lambda = 100 \tev$.

\subsection{Bileptons}

The peculiar initial state quantum numbers of $\emem$ colliders make
them uniquely suited to exploring a variety of exotic phenomena.
Chief among these are bileptons, particles with lepton number $L=\pm
2$.  Such particles appear, for example, in models where the SU(2)$_L$
gauge group is extended to SU(3)~\cite{Frampton}, and the Lagrangian
contains the terms
\begin{equation}
{\cal L} \supset \left(\begin{array}{ccc} \ell^-& \nu & \ell^+
\end{array}\right)_L^* 
\left( \begin{array}{ccc}
 &  & Y^{--} \\
 & & Y^- \\
Y^{++} & Y^+ & \end{array} \right) 
\left( \begin{array}{c} \ell^- \\ \nu \\ \ell^+ \end{array} \right)_L
 \ ,
\end{equation}
where $Y$ are new gauge bosons.  $Y^{--}$ may then be produced as an
$s$-channel resonance at $\emem$ colliders, mediating background-free
events like $\emem \to Y^{--} \to \mu^- \mu^-$.  Clearly the $\emem$
collider is ideally suited to such studies.

Bileptons are also obtained in models with extended Higgs sectors that
contain doubly charged Higgs bosons $H^{--}$.  The potential of
$\emem$ colliders to probe resonances and other phenomena in these
models has been reviewed by Gunion~\cite{Gunion}.

\subsection{Supersymmetry}

Supersymmetry would appear at first sight to be a perfect example of
new physics that is difficult to explore at $\emem$ colliders.
Indeed, the dictatorship of leptons forbids the production of most
superpartners: $e^- e^- \not\to \chi^0 \chi^0, \chi^- \chi^-,
\tilde{q} \tilde{q}^*, \tilde{\nu} \tilde{\nu}^*$.  However, all
supersymmetric models contain Majorana fermions that couple to
electrons: the gauginos $\tilde{B}$ and $\tilde{W}$.  As was noted
long ago by Keung and Littenberg~\cite{KL}, these mediate selectron
pair production through the process shown in Fig.~\ref{fig:selectron}.

\begin{figure}[t]
\centerline{\epsfig{file=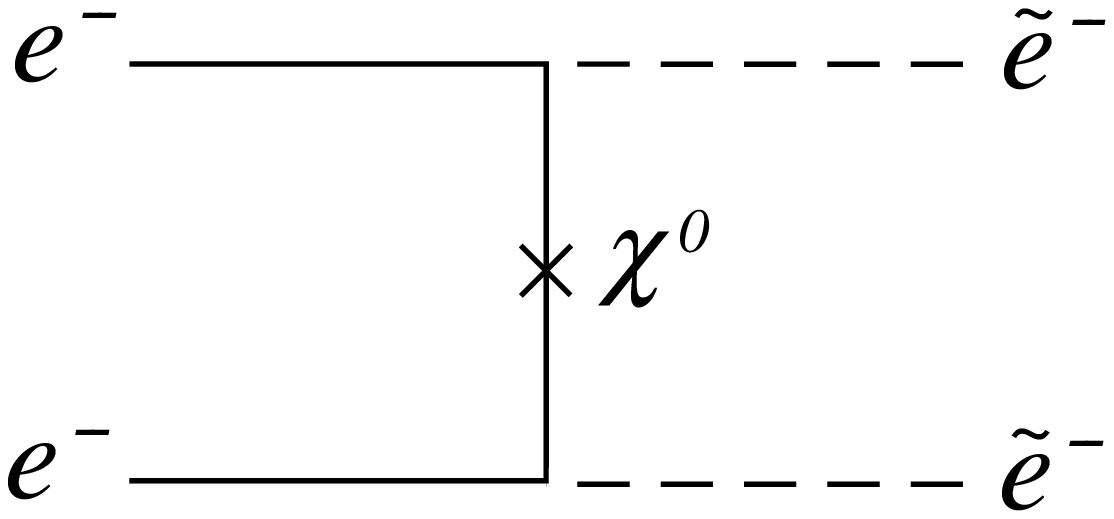,width=0.57\textwidth}}
\vspace*{0.1in} 
\fcaption{Selectron pair production $e^- e^- \to \selectron^-
\selectron^-$, mediated by $t$-channel Majorana neutralino
exchange.}
\label{fig:selectron}
\end{figure}

Although supersymmetry at $\emem$ colliders is limited to slepton pair
production, studies of slepton masses, mixings, and couplings can
yield a great deal of information and provide excellent examples of
how the properties of $\emem$ colliders may be exploited.  Let us
consider them in turn.

\subsubsection{Masses}

Masses at linear colliders are most accurately determined through
either kinematic endpoints~\cite{endpoints} or threshold
scans~\cite{threshold}. In a recent study of $\epem$ colliders, Martyn
and Blair have considered both possibilities~\cite{MB}.  For the pair
production of fermions such as charginos (see Fig.~\ref{fig:mass}a),
the cross section at threshold rises as $\beta$, the velocity of the
produced particles.  Threshold scans are then highly effective, and
typical accuracies achieved are $\Delta m \sim 10 - 100 \mev$.  For
the pair production of identical scalars, the cross section rises as
$\beta^3$ at threshold, and so threshold studies, though possible with
very large luminosities~\cite{MB}, are much less effective.  Instead
one turns to kinematic endpoints (see Fig.~\ref{fig:mass}b), where
mass measurements typically yield $\Delta m \sim 0.1 - 1 \gev$.

\begin{figure}[t]
\epsfig{file=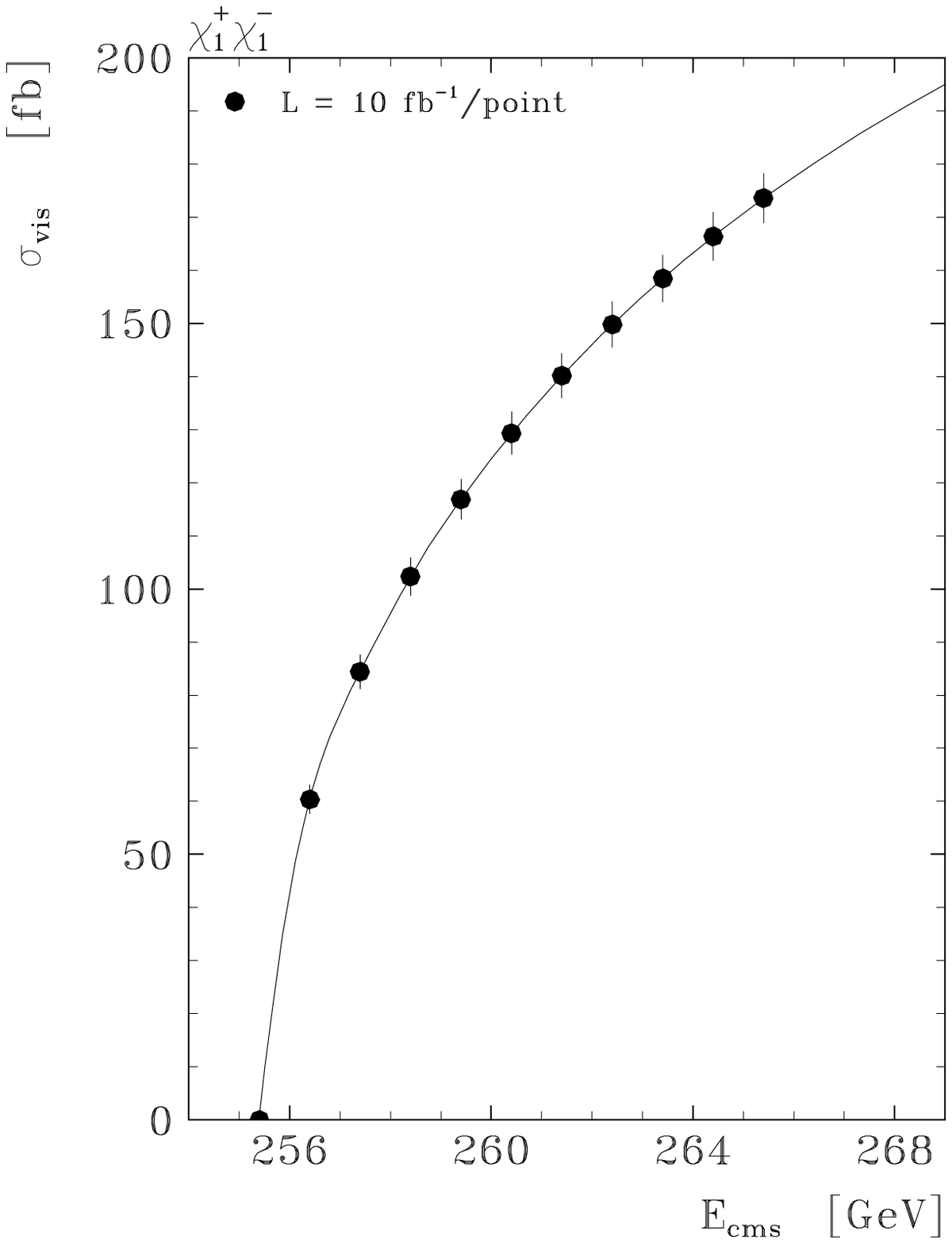,width=0.28\textwidth}
\hfill
\epsfig{file=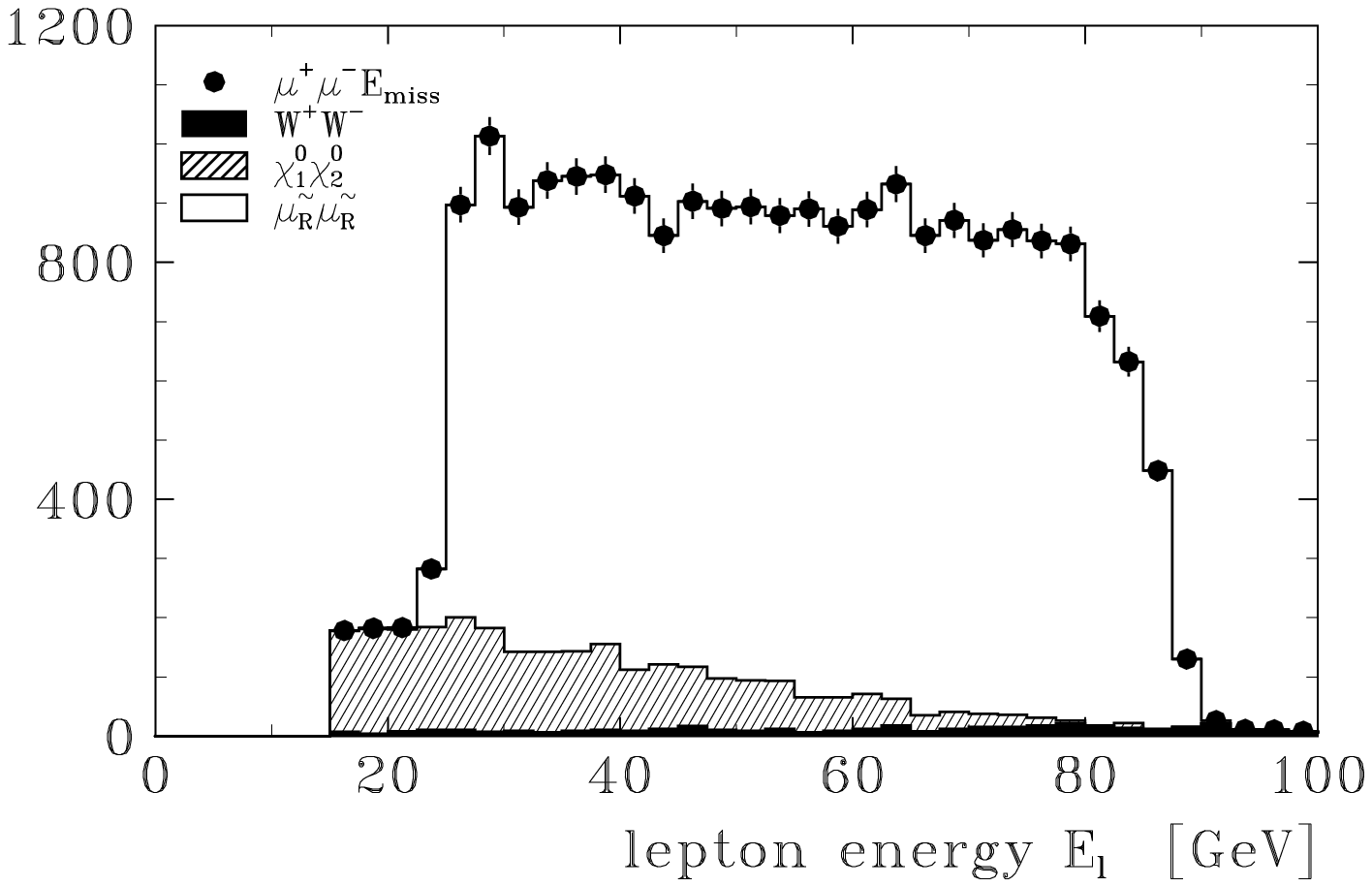,width=0.56\textwidth}
\vspace*{0.1in} 
\fcaption{Mass determination at $\epem$ colliders (a) for charginos via
threshold scanning, and (b) for smuons via kinematic endpoints. (From the
study of Martyn and Blair~\cite{MB}.)}
\label{fig:mass}
\end{figure}

At $\emem$ colliders, however, the same-helicity selectron pair
production cross section has a $\beta$ dependence at threshold. This
is easily understood: the initial state in $e^-_R e^-_R \to
\selectron^-_R \selectron^-_R$ has angular momentum $J=0$, and so the
selectrons may be produced in the $S$ wave state.  The unique quantum
numbers of $\emem$ colliders therefore effectively convert a kinematic
endpoint measurement into a threshold measurement (see
Fig.~\ref{fig:compare}), and extremely accurate scalar mass
measurements are possible with minimal cost in luminosity.
Incidentally, the full arsenal of linear collider modes allows one to
extend this mass measurement to the rest of the first generation
sleptons through a series of $\beta$ threshold scans: $e^-e^- \to
\tilde{e}^-_R \tilde{e}^-_R$ yields $m_{\tilde{e}_R}$; $e^+e^- \to
\tilde{e}^{\pm}_R \tilde{e}^{\mp}_L$ yields $m_{\tilde{e}_L}$; $e^+e^-
\to \chi^+ \chi^-$ yields $m_{\chi^{\pm}}$; and $e^- \gamma \to
\tilde{\nu}_e \chi^-$ yields $m_{\tilde{\nu}_e}$.

\begin{figure}[t]
\centerline{\epsfig{file=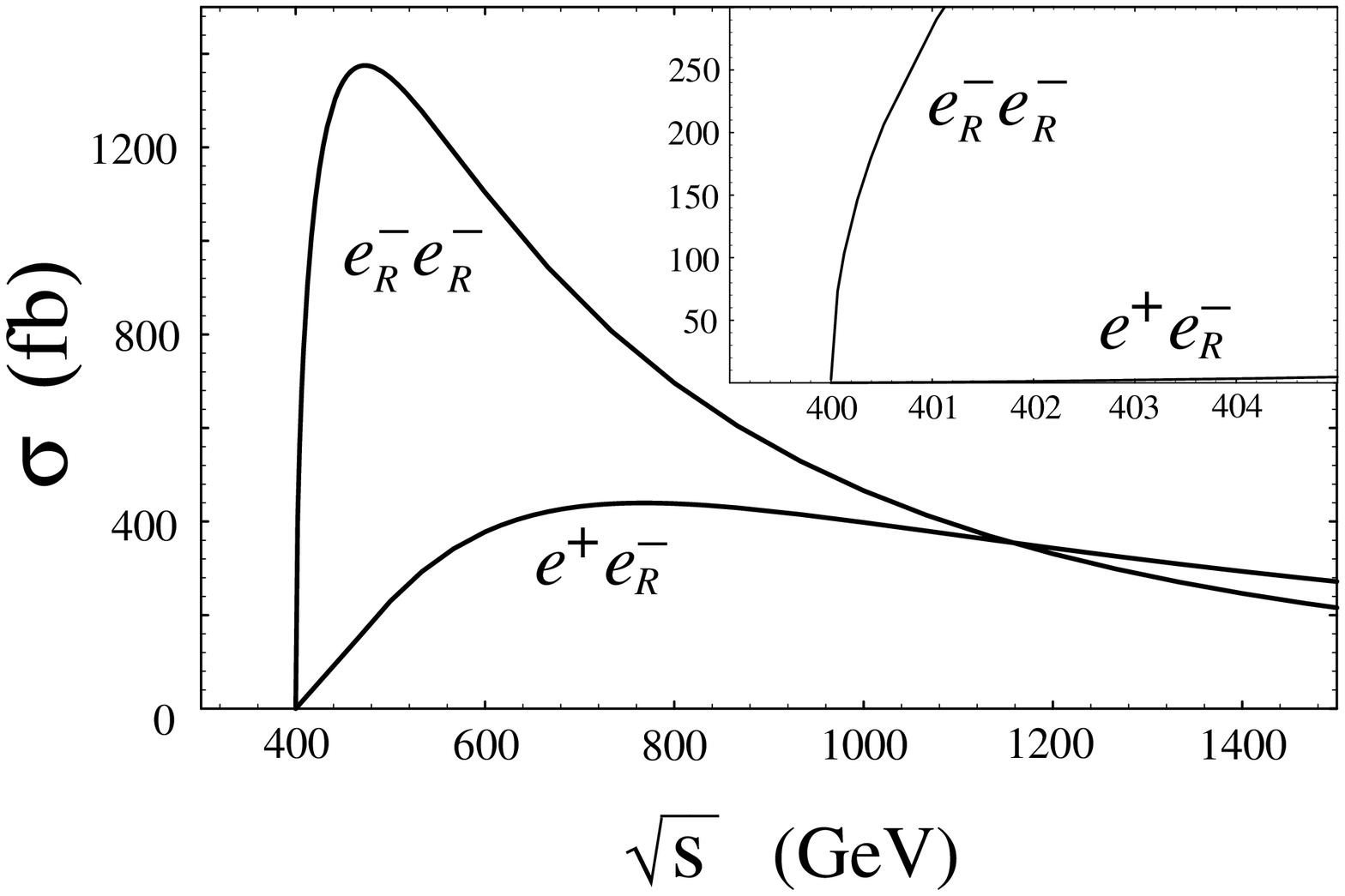,width=0.62\textwidth}}
\fcaption{Cross sections for selectron pair production at $\epem$ and
$\emem$ colliders: $\sigma(e^-_R e^-_R \to \selectron^-_R
\selectron^-_R)$ and $\sigma(e^+e^-_R \to \selectron^+_R
\selectron^-_R)$~\cite{F97}.  The inset is a magnified view near
threshold. Initial state radiation, beamstrahlung, and finite width
effects are not included.}
\label{fig:compare}
\end{figure}

\subsubsection{Mixings}

Now that neutrinos are known to mix, lepton flavor is no longer a
sacred symmetry, and there is every reason to expect that sleptons
also have generational mixings.  Such mixing leads to decays
$\tilde{e} \to \mu \chi^0$ and may be searched for at either $\epem$
or $\emem$ colliders.

At $\epem$ colliders, the signal is $e^+e^- \to e^{\pm} \mu^{\mp}
\chi^0 \chi^0$.  The backgrounds are
\begin{eqnarray}
e^+e^- \to W^+ W^- &\quad& {\rm single}\ e^-_R\ 
{\rm polarization} \nonumber \\ 
e^+e^- \to \nu \nu W^+ W^- &\quad& {\rm single}\ e^-_R \ 
{\rm polarization} \nonumber \\
e^+e^- \to e^{\pm} \nu W^{\mp} && \nonumber \\ 
\gamma \gamma \to W^+ W^- &&
\end{eqnarray}
The first two backgrounds may be reduced by beam polarization, as
indicated. However, the last two are irreducible.

In the $\emem$ case, the signal is $e^-e^- \to e^- \mu^- \chi^0
\chi^0$.  Possible backgrounds are
\begin{eqnarray}
e^-e^- \to W^- W^- &\quad& {\rm forbidden\ by\ total }\ 
L\ {\rm number} \nonumber \\
e^-e^- \to \nu \nu W^- W^- &\quad& {\rm single}\ e^-_R \ 
{\rm polarization} \nonumber \\
e^-e^- \to e^- \nu W^- &\quad& {\rm double}\ e^-_R\ 
{\rm polarization} \nonumber \\
\gamma \gamma \to W^+ W^- &\quad& {\rm same\ sign\ leptons}
\end{eqnarray}
In this case, all backgrounds may be eliminated, in the limit of
perfect beam polarization.  As a result, the sensitivity of $\emem$
colliders to slepton flavor violation is much greater than at $\epem$
colliders, and is also much more sensitive than current and near
future low energy experiments~\cite{ACFH}.

\subsubsection{Couplings}

The excellent properties of $\emem$ colliders for exploring selectron
production also make possible extremely precise determinations of
selectron gauge couplings. Denoting the $e \tilde{e} \tilde{B}$ and $e
e B^{\mu}$ couplings by $h$ and $g$ respectively, it is possible to
verify $h/g=1$ to well below the percent level~\cite{couplings}.  This
then provides a quantitative check of supersymmetry and allows one to
verify that the selectron is in fact the superpartner of the electron.

This measurement takes on additional importance if one notes that the
relation $h/g=1$ is modified by heavy superpartners, and the deviation
grows logarithmically with the superpartner mass
scale~\cite{superoblique} --- that is, $h/g-1$ is a non-decoupling
observable that receives contributions from arbitrarily heavy
superpartners!  Superheavy superpartners are phenomenologically
attractive in many ways and may be present in a wide variety of models
without sacrificing naturalness~\cite{superheavy}.  A measurement of
$h/g$ then provides one of the few probes of kinematically
inaccessible superpartners and may help set the scale for far future
colliders.

\section{Conclusions}

I have briefly reviewed the merits of $\emem$ colliders.  The ability
to highly polarize both beams and the unique quantum numbers of the
initial state provide novel opportunities to study new physics.

A few illustrative examples were presented --- of course, there are
many more possibilities.  I have taken the liberty of grossly
oversimplifying matters by summarizing each theoretical talk at this
conference with a single Feynman diagram (or, in exceptional cases,
two). These are presented in Fig.~\ref{fig:all}. It is evident that
the topics covered span a broad range, and include top quarks, Higgs
bosons, extra gauge bosons, Majorana neutrino masses, strong $WW$
scattering, and processes involving external and internal graviton
states.  Of course, to judge the effectiveness of $\emem$ colliders,
it is important not just that $\emem$ colliders are sensitive to such
physics, but that $\emem$ colliders provide probes at least as
effective as or complementary to those available at the LHC, $\epem$
colliders, and low energy experiments, with reasonable experimental
assumptions.  Such important considerations will be addressed by the
following speakers.

It is clear that in some scenarios, the unique properties of $\emem$
colliders will provide additional information through new channels and
observables.  While the specific scenario realized in nature is yet to
be determined, given the exciting and possibly confusing era we are
about to enter, such additional tools may prove extremely valuable in
elucidating the physics of the weak scale and beyond.

\begin{figure}[t]
\epsfig{file=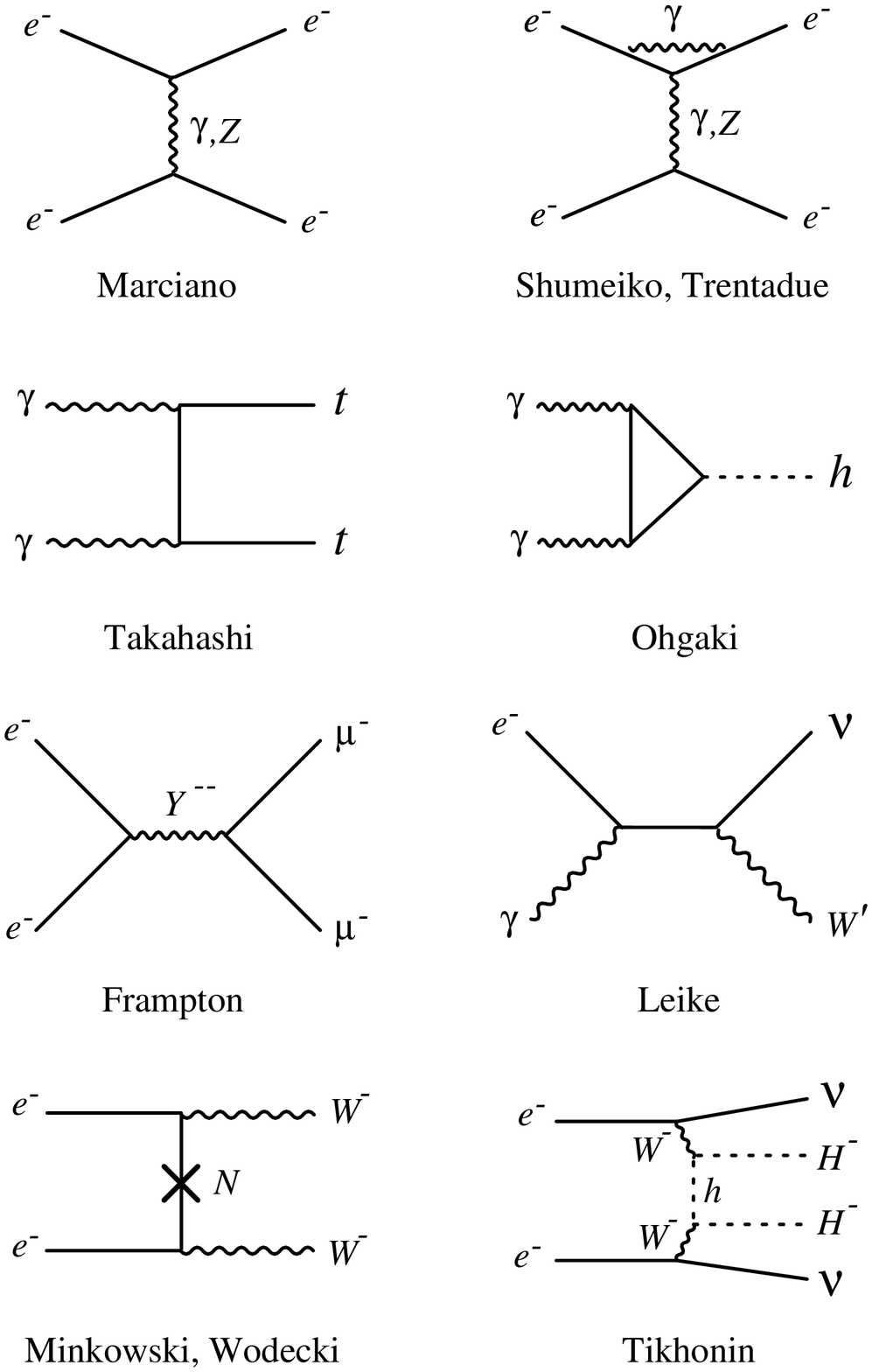,width=0.49\textwidth}
\hfill
\epsfig{file=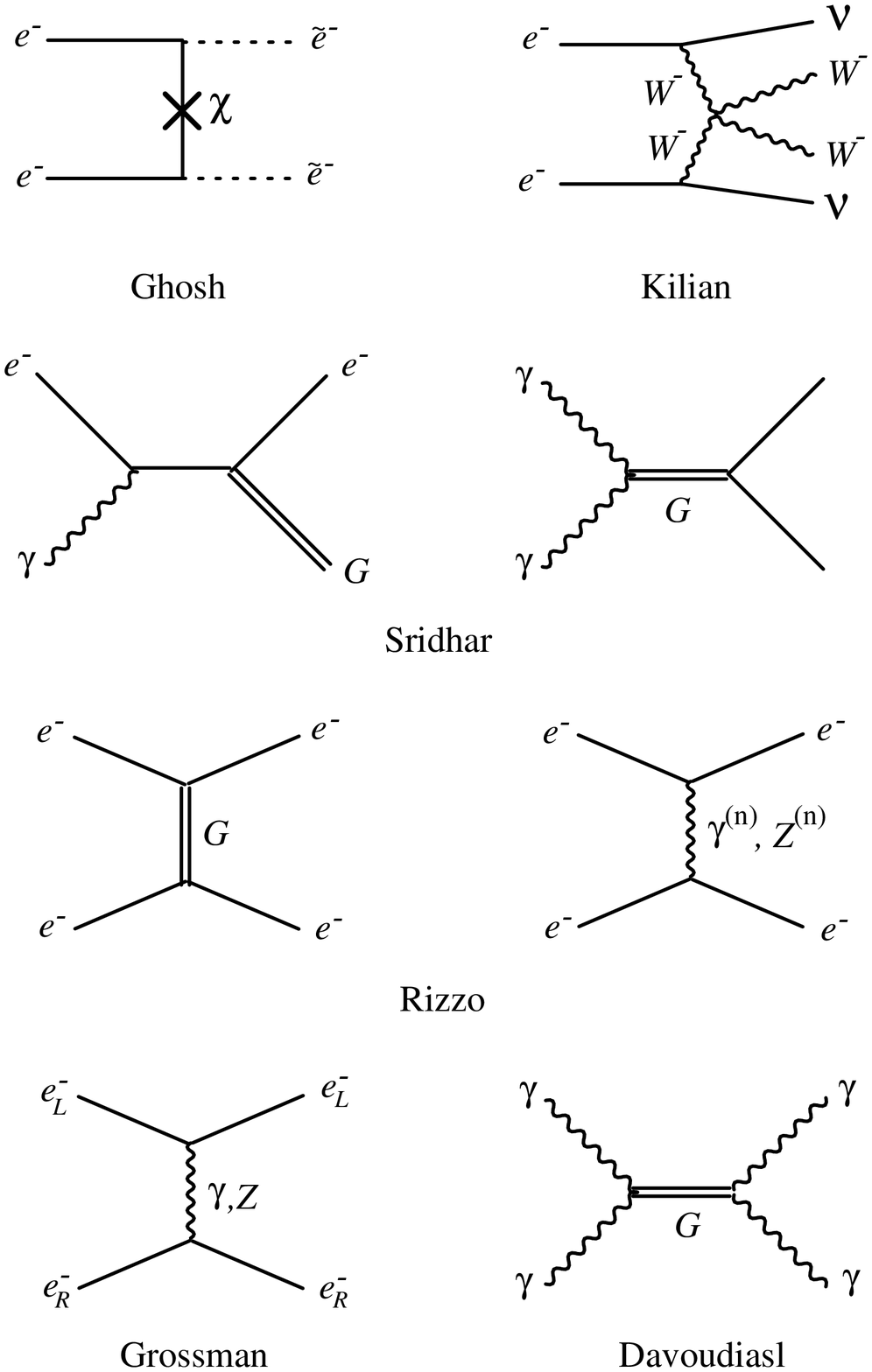,width=0.49\textwidth}
\vspace*{0.1in} 
\fcaption{Feynman diagram summary of talks presented at this
conference. }
\label{fig:all}
\end{figure}

\nonumsection{Acknowledgments}

I am grateful to the organizers, especially C.~Heusch and N.~Rogers,
for a stimulating and enjoyable conference.  This work was supported
in part by the Department of Energy under contract DE--FG02--90ER40542
and through the generosity of Frank and Peggy Taplin.
 
\nonumsection{References}

\end{document}